\documentclass[12pt]{article}
\usepackage{jheppub}
\usepackage{MnSymbol}
\def\CA{{\cal A}}
\def\CB{{\cal B}}
\def\CD{{\cal D}}
\def\CM{{\cal M}}
\def\CN{{\cal N}}
\def\CS{{\cal S}}
\def\CT{{\cal T}}
\def\CI{{\cal I}}
\def\bZ{{\mathbb Z}}

\title{Domain Walls for Two-Dimensional Renormalization Group Flows}

\author[1]{Davide Gaiotto}

\affiliation[1]{Institute for Advanced Study, Einstein Dr., Princeton, NJ 08540, USA}

\abstract{Renormalization Group domain walls are
natural conformal interfaces between two CFTs related by an RG flow. The RG domain wall 
gives an exact relation between the operators in the UV and IR CFTs. 
We propose an explicit algebraic construction of the RG domain wall 
between consecutive Virasoro minimal models in two dimensions. 
Our proposal passes a stringent test: it reproduces in detail the leading order mixing 
of UV operators computed in the conformal perturbation theory literature. 
The algebraic construction  
can be applied to a variety of known RG flows in two dimensions. 
}

\begin{document}

\maketitle

\section{Introduction}
Distinct conformal field theories can be often coupled by conformal invariant interfaces (see e.g. \cite{Bachas:2001vj} and references therein). 
A conformal interface gives a map between observables in the two theories. 
For example, in radial quantization, a spherical interface gives a map between the Hilbert spaces of the two theories on the sphere, i.e. between local operators. Such maps may be trivial: the interface may just be a totally reflecting interface, i.e.  the product of independent boundary conditions for the two theories. But there are situations where a conformal interface may encode in a compact fashion an important relation between two theories
\cite{Douglas:2010ic}.

In supersymmetric, four-dimensional examples of SCFTs with exactly marginal deformations, 
such as $\CN=4$ SYM or $\CN=2$ gauge theories, there are notable examples of such interfaces. 
Janus interfaces encode the parallel transport of operators of the theory along the parameter space of exactly marginal deformations \cite{D'Hoker:2006uv}.
S-duality walls encode the combination of the parallel transport with an S-duality operation,  
and give a map between weakly-coupled descriptions of the same theory valid 
in different corners of moduli space \cite{Gaiotto:2008ak}.

There is another situation where a map between the operators of two CFTs arises: renormalization group flows. 
Consider a situation where a CFT $\CT_{UV}$ can be perturbed by a relevant operator $\phi$
to give an RG flow to a second CFT $\CT_{IR}$. Then the operators of $\CT_{UV}$
are mixed and renormalized to give operators in $\CT_{IR}$. The corresponding map can be made very explicit in 
examples where conformal perturbation theory is valid \cite{Zamolodchikov:1987ti}. 

\begin{figure}[htb]
\centering
\includegraphics[width=5in]{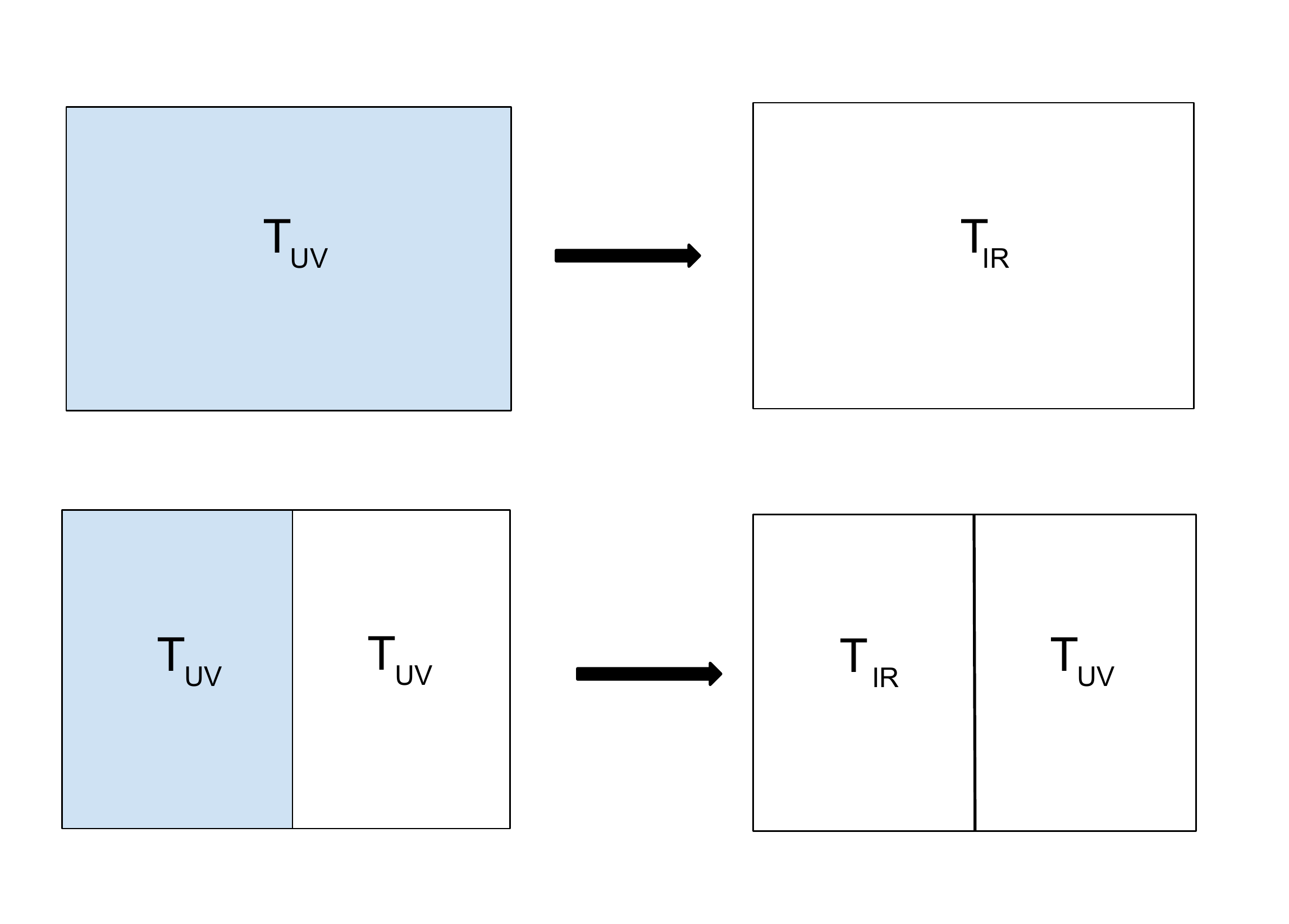}
\caption{An RG flow between theories $\CT_{UV}$ and $\CT_{IR}$ is initiated by perturbing $\CT_{UV}$ by a relevant operator integrated over the whole space. 
If the relevant operator is integrated over a half-space only, the RG flow will produce a RG interface between $\CT_{UV}$ and $\CT_{IR}$}
\label{fig:one}
\end{figure}

It is natural to wonder if a RG conformal interface may exist, which encodes such correspondence between observables in 
$\CT_{UV}$ and $\CT_{IR}$. There is an obvious candidate \cite{Fredenhagen:2005an,Brunner:2007ur}. Start with $\CT_{UV}$, and perturb it by the integral of $\phi$ over a half space $x_1<0$ only. 
Flowing to the far infrared, the result will be a conformal interface between $\CT_{UV}$ and $\CT_{IR}$, which we will refer to as RG domain wall, or RG interface. See Figure \ref{fig:one}

In a general setup, this definition is a bit ambiguous, even if we specify the precise choice of RG flow for the bulk theory. 
As the perturbation by the integral of $\phi$ on a half-space breaks more symmetries than the perturbation by the integral of $\phi$ over the whole space, 
renormalization may turn on other relevant operators which preserve the reduced symmetries of the setup. For example, 
it can turn on operators integrated over the interface only.  Thus a precise definition of the RG domain wall would require 
one to specify a RG flow in such extended space of couplings. The definition can be sharpened if the RG flow preserves some additional symmetry, such as supersymmetry. 

A beautiful example of supersymmetric RG domain walls in two-dimensional ${\cal N}=2$ supersymmetric minimal models is given in \cite{Brunner:2007ur}. The RG domain walls are described in the language of matrix factorizations in a Landau-Ginzburg orbifold description of the minimal models. The description is sufficiently powerful to allow the computation of fusion between the RG domain walls and supersymmetric boundary conditions. The results reproduce the expected behavior of supersymmetric boundary conditions under the RG flow. Notice that in the mirror Landau-Ginzburg description, there is little difference between RG domain walls and 
Janus domain walls as defined e.g. in \cite{Brunner:2008fa,Gaiotto:2009fs}.

In this note we will focus on less-supersymmetric two-dimensional examples of RG domain walls,
for which $\CT_{UV}$ and $\CT_{IR}$ are both rational conformal field theories,
related by a well understood RG flow which can be described in conformal perturbation theory. 
The RG domain wall should be a conformal interface 
which is perturbatively close to the trivial identity interface, and should only pair up primary fields of $\CT_{UV}$ and $\CT_{IR}$ 
which are related by the perturbative RG flow. These conditions taken together are rather constraining, 
and one may hope they can identify uniquely a candidate RG domain wall. 

\begin{figure}[htb]
\centering
\includegraphics[width=5in]{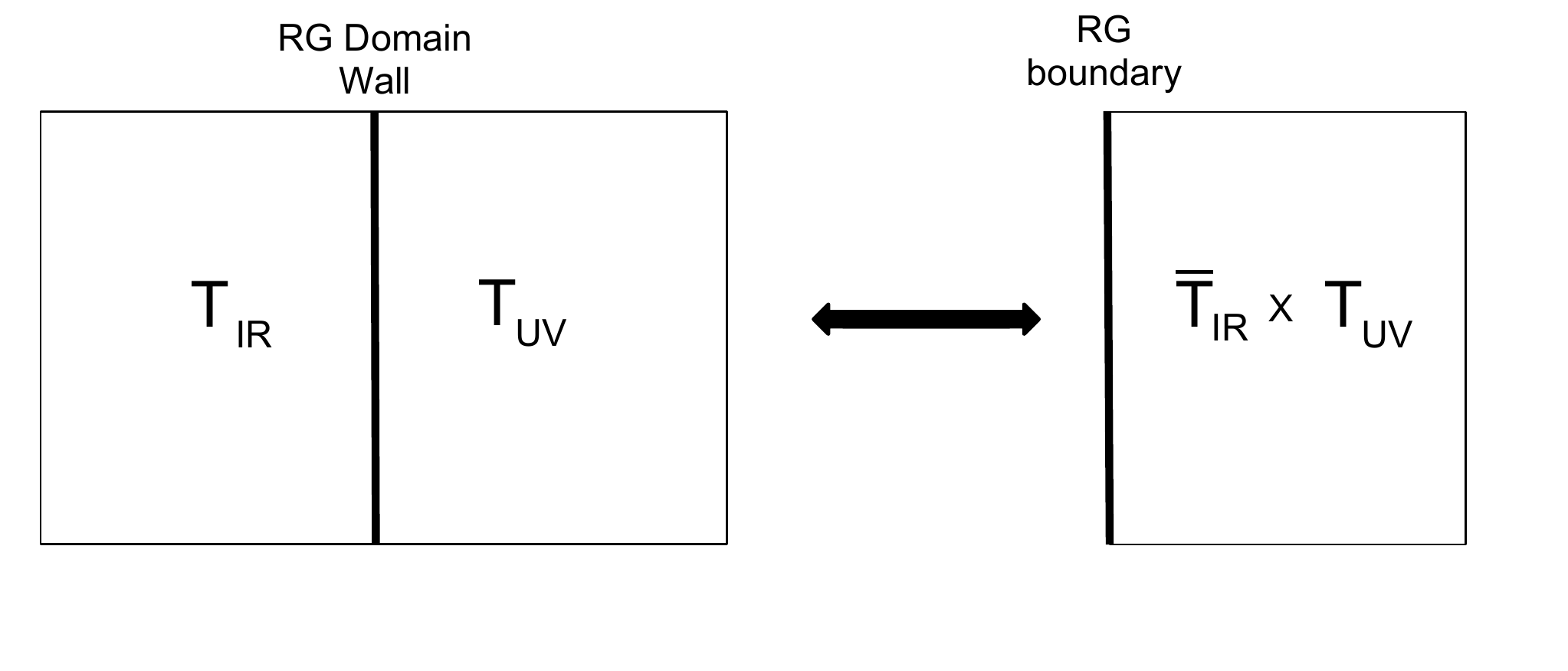}
\caption{The folding trick: a conformal interface between theories $\CT_{UV}$ and $\CT_{IR}$ can be reinterpreted as a conformal boundary condition in the product theory $\CT_{UV} \times \CT_{IR}$. Holomorphic and anti-holomorphic quantities in
$\CT_{IR}$ are exchanged. For example, the gluing condition for the stress tensor for a conformal interface $T_{UV} - \bar T_{UV}=T_{IR} - \bar T_{IR}$ becomes the gluing condition for a conformal boundary 
$T_{UV} + T_{IR}= \bar T_{UV} + \bar T_{IR}$  }
\label{fig:two}
\end{figure}

This program meets a simple obstruction: general conformal interfaces between RCFTs are not rational, and do not admit a simple Cardy algebraic description. 
Conformal interfaces can be described as conformal boundary conditions for the product $\CT_{UV} \times \CT_{IR}$ 
by the standard folding trick (see Figure \ref{fig:two}). Boundary conditions which preserve the stress-tensors of the two factors independently,  $T_{UV} = \bar T_{UV}$ and $T_{IR} = \bar T_{IR}$
corresponds to totally reflective interfaces. A general conformal interface only satisfies $T_{UV} - \bar T_{UV}=T_{IR} - \bar T_{IR}$, and corresponds to a boundary condition which only preserves the 
total stress tensor of the product theory. 

Rational boundary conditions in the product theory $\CT_{UV} \times \CT_{IR}$ 
must glue the chiral algebra of the product theory $\CA_{UV} \times \CA_{IR}$ to the anti-chiral algebra by some automorphism. 
Generically, there is no automorphism which acts non-trivially on the individual stress tensors, and all rational boundary conditions 
of the product theory correspond to totally reflective interfaces. 
 
The expectation for a RG interface is to be as far as possible from a totally reflective interface. 
In the perturbative setup, the RG interface should be perturbatively close to the identity interface, which is totally transmissive.
Thus the RG boundary condition, i.e. the boundary condition in $\CT_{UV} \times \CT_{IR}$ which corresponds to the 
RG domain wall, cannot be rational unless $\CT_{UV} \times \CT_{IR}$ has a hidden symmetry. This is not the case in the examples of
RG flows we consider. 

Sometimes, it is possible to give an algebraic definition of non-trivial interfaces by relating the interface to a rational boundary condition in a {\it different} 2d CFT.
Beautiful examples of such construction can be found in \cite{Quella:2002ct,Quella:2003kd} and \cite{Quella:2006de}, which also quantifies the idea that an interface can be more or less reflective or transmissive. 
Although the specific construction of these references does not apply to RG domain walls, in this paper we identify a class of RG flows for which an alternative construction exists. 
This allows us to give an algebraic construction of many RG interfaces in terms of rational boundary condition for a different 2d CFT $\CT_\CB$.
 
Our main example will be a pair of consecutive unitary minimal models, but 
our strategy applies to a much larger class of examples, such as the cosets 
\begin{equation}
\CT_{UV} = \frac{\hat g_l \times \hat g_m}{\hat g_{l+m}} \qquad \qquad m > l
\end{equation}
which can be deformed by the $\phi_{1,1}^{\mathrm{Adj}}$ coset  field \cite{Ravanini:1992fs} to flow to 
\begin{equation}
\CT_{IR} = \frac{\hat g_l \times \hat g_{m-l}}{\hat g_{m}} \qquad \qquad m > l
\end{equation}

The crucial observation is that in these models, the relevant perturbation $\phi$ 
commutes with a set of topological interfaces in $\CT_{UV}$ \cite{Fredenhagen:2009tn}.
Topological interfaces in $\CT_{UV}$ which commute with $\phi$ 
are not renormalized by the RG flow, and survive as topological interfaces in $\CT_{IR}$. 
We will argue that the RG domain wall must simply intertwine 
such topological interfaces in $\CT_{UV}$ and their image in $\CT_{IR}$. 

Starting from this observation, we will argue that the RG domain wall is closely related to a rational brane 
in the theory 
\begin{equation}
\CT_\CB =\frac{\hat g_l \times \hat g_l \times \hat g_{m-l}}{\hat g_{l+m}}
\end{equation}
The theory $\CT_\CB$ has a a large chiral algebra $\CB$ which contains $\CA_{UV} \times \CA_{IR}$ as a subalgebra. 
Indeed, $\CT_\CB$ can be thought of as a non-diagonal modular invariant for $\CA_{UV} \times \CA_{IR}$, and there is a convenient canonical 
topological interface $\CI_1$ between $\CT_\CB$ and $\CT_{UV} \times \CT_{IR}$.

\begin{figure}[htb]
\centering
\includegraphics[width=5in]{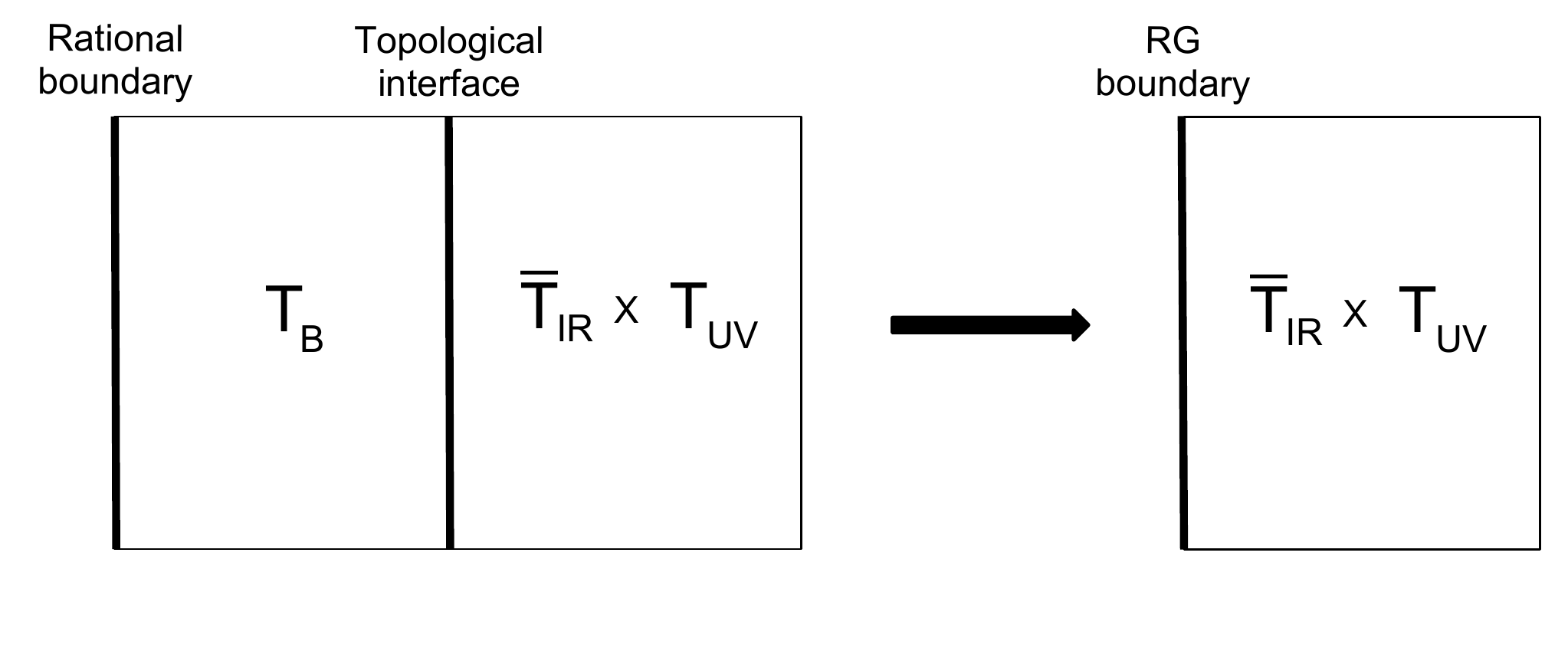}
\caption{We produce a non-rational candidate for the RG boundary condition by colliding a rational topological interface between the product theory $\CT_{IR} \times \CT_{UV}$ and the auxiliary theory 
$\CT_\CB$ and a rational, twisted boundary condition in $\CT_\CB$.}
\label{fig:three}
\end{figure}

The chiral algebra $\CB$ has an obvious $Z_2$ symmetry under exchange of the first two factors in the numerator. 
We propose that the RG boundary condition is the image under the action of $\CI_1$ of 
a specific $Z_2$-twisted brane in $\CT_\CB$. See Figure \ref{fig:three}. This gives a completely algebraic description of 
our candidate RG domain wall. For minimal models, we formulate a simple algorithm to 
compute the $Z_2$-twisted Ishibashi states of $\CB$. 

Our calculation only involves representation theory of the chiral algebra $\CB$, 
and outputs the pairing between a specific $UV$ operator $O^{UV}$ and a specific $IR$ operator $O^{IR}$,
in the form of a disk one-point function for the candidate RG brane
\begin{equation}
\langle \bar O^{IR} O^{UV} | RG \rangle \equiv \langle O^{IR}|RG| O^{UV} \rangle
\end{equation}

We interpret this pairing of local operators in $\CT_{IR}$ and $\CT_{UV}$ as describing the exact mixing of UV operators under the RG flow to give IR operators.
In order to test this interpretation  
we compare it with explicit calculations done in the literature on conformal perturbation theory \cite{Zamolodchikov:1987ti}. 
The results match perfectly. It is worth observing that the two calculations 
are not related in any obvious way. The intricate mixing coefficients are computed from 
apparently different subsets of information about the two-dimensional CFTs.

The conformal perturbation theory calculation involves the diagonalization of the leading order matrix of anomalous dimensions, 
which is computed by the three-point functions 
\begin{equation}
C_{ij\phi} = \langle O^{UV}_i O^{UV}_j \phi \rangle
\end{equation}
for several UV operators $O^{UV}_a$ of similar conformal dimension. 
The various IR operators correspond to the eigenvectors of the anomalous dimension matrix.

Our algebraic proposal reproduces one-by-one, independently, 
the individual matrix elements of the eigenvectors.
We take this match as strong evidence that our strategy to identify the RG domain wall is correct.

It would be nice to test the correspondence further, 
for example compare our results to more detailed calculations in the canonical example of minimal models, and in the more general class of models to which our proposal apply. 
It should be also possible to study the collision between the candidate RG domain wall and branes in $\CT^{UV}$, and compare the result with the expected behavior of
boundary conditions under RG flow. 

The RG flows we consider are known to lead to integrable theories at intermediate 
scales between the UV and the far IR \cite{Zamolodchikov:1987jf}.  This fact played no role in our calculation. It would be interesting to see if some interplay is possible between 
the RG domain wall and integrability. In principle, we expect to have RG domain walls across which the RG scale jumps by a finite amount, thus relating 
the integrable theory at two points along the RG flow. But it is unclear if such interfaces might be in some sense integrable.
 
Finally, the class of cosets we discuss includes the $W_N$ minimal models, $g = su(N)$ and $l=1$, which have a conjectural holographic dual in the 
'tHooft limit of large $k$ and $N$ with fixed $N/k$ \cite{Gaberdiel:2010pz}.\footnote{We thank Shlomo Razamat for this observation} The RG flow we consider has a neat holographic interpretation 
as a change in boundary conditions for a bulk scalar field. It would be interesting to look for a holographic dual description of the RG domain wall,
possibly as a boundary-changing interface for the same bulk scalar field. It would also be interesting to match the behavior of two-point functions and topological defects 
in the bulk and boundary theories. 
 
 \section{RG domain wall between Virasoro minimal models} 
 \subsection{Conformal perturbation theory}
The best known RG flow in two dimensions relates a consecutive pair of unitary minimal models \cite{Zamolodchikov:1987ti}
$\CT_{UV}=\CM_{p+1,p}$ and $\CT_{IR} = \CM_{p,p-1}$.
Remember that the minimal model $\CM_{k+3,k+2}$ is a RCFT for the Virasoro algebra, with central charge
\begin{equation}
c = 1-\frac{6}{(k+2)(k+3)}
\end{equation}
and chiral primaries $\phi_{r,s}$ of conformal dimension 
\begin{equation}
h_{r,s} = \frac{(r-s)^2}{4} + \frac{r^2-1}{4(k+2)} - \frac{s^2-1}{4(k+3)} \end{equation}
in the ranges $1 \leq r \leq k+2 $ and $1 \leq s \leq k+3$ with an identification 
\begin{equation}
\phi_{k+2-r,k+3-s} = \phi_{r,s}
\end{equation} 
In the following, we will refer to the RCFT defined by the diagonal modular invariant for the minimal model. 

In the large $k$ regime, the fields $\phi_{r,r}$ have conformal dimension close to $0$, $\phi_{r,r \pm 1}$ close to $1/4$, 
$\phi_{r,r \pm 2}$ close to $1$. In particular, $\phi_{1,3}$ is the least relevant field for all $k$. 
The minimal model $\CM_{p+1,p}$ perturbed by $\phi^{UV}_{1,3}$ will flow to  $\CM_{p,p-1}$. 
The flow will arrive to  $\CM_{p,p-1}$ along the least irrelevant direction, corresponding to $\phi^{IR}_{3,1}$.
Here and in the following we distinguish the fields in the UV and IR by the corresponding superscripts.

The field $\phi_{1,3}$ has particularly simple fusion rules with other fields $\phi_{r,s}$: it only changes the label $s$ by an even amount. 
This implies that in conformal perturbation theory no other relevant perturbations need to be turned on, as the other $\phi_{1,s}$ for odd $s$ 
are all irrelevant. 
Furthermore, the mixing among fields in the UV theory is very constrained. The perturbative analysis shows that 
the Verma modules of $\phi^{UV}_{r,s}$ with the same $r$ and all possible $s$ with the same parity 
will mix together to give the Verma 
modules of $\phi^{IR}_{t,r}$ for all $t$ with the same parity as $s$. Schematically,
\begin{equation}
\sum_s^{\mathrm{even}} \left[ \phi^{UV}_{r,s} \right] \to \sum_t^{\mathrm{even}} \left[ \phi^{IR}_{t,r} \right]  \qquad \qquad \sum_s^{\mathrm{odd}} \left[ \phi^{UV}_{r,s} \right] \to \sum_t^{\mathrm{odd}} \left[ \phi^{IR}_{t,r} \right]  
\end{equation} 

In the perturbative regime, $k \gg 1$, only operators whose dimensions differ by order $k^{-1}$ mix at the leading order. 
One can compute the matrix of anomalous dimensions $\gamma_{ij}$ at order $k^{-1}$ from three-point functions involving
$\phi_{1,3}$:
\begin{equation}
\gamma_{ij} = h_i \delta_{ij} + \frac{\sqrt{3}}{k} C_{i,j,(1,3)} 
\end{equation}
Here $2 \pi g_* = \frac{\sqrt{3}}{k}$ is the critical value of the coupling. 

 The eigenvectors of $\gamma_{ij}$ describe the leading order mixing 
of UV operators to give IR operators. The eigenvectors are matched to the known IR operators 
by looking at the conformal dimensions 
at order $k^{-1}$ which are computed from the eigenvalues of  $\gamma_{ij}$.

The mixing pattern is easy to describe for $r$ not too large compared to $k$. 
The operator $\phi^{UV}_{r,r}$ has the smallest possible dimension, 
of order $k^{-2}$, and does not mix: it flows directly to $\phi^{IR}_{r,r}$.

The operators $\phi^{UV}_{r,r \pm 1}$ have dimensions which is close to $1/4$. 
They mix at the leading order, and the eigenvectors of the matrix of anomalous dimensions given in \cite{Zamolodchikov:1987ti}
are
\begin{align}
\phi^{IR}_{r-1,r} &= \frac{\sqrt{r^2-1}}{r} \phi^{UV}_{r,r+1} - \frac{1}{r} \phi^{UV}_{r,r-1} \cr
\phi^{IR}_{r+1,r} &= \frac{1}{r}  \phi^{UV}_{r,r+1} + \frac{\sqrt{r^2-1}}{r} \phi^{UV}_{r,r-1} \cr
\end{align}
We fixed the normalization so that the operators have norm one, and the sign so that the mixing 
becomes trivial as $r$ becomes large, as the dimensions of the UV operators differ more and more. 

The operators $\phi^{UV}_{r,r \pm 2}$ and $\partial \bar \partial \phi^{UV}_{r,r}$ have dimensions close to $1$, 
and mix at the leading order as (again, we simply diagonalized the matrix of anomalous dimensions from \cite{Zamolodchikov:1987ti} and fixed the normalization properly)
\begin{align}
\phi^{IR}_{r+2,r} &= \frac{2}{r(r+1)} \phi^{UV}_{r,r+2}+ \frac{2}{r+1} \sqrt{\frac{r+2}{r}} (2 h^{UV}_{r,r})^{-1} \partial \bar \partial \phi^{UV}_{r,r} + \frac{ \sqrt{r^2-4}}{r} \phi^{UV}_{r,r-2} \cr
\phi^{IR}_{r-2,r} &= \frac{\sqrt{r^2-4}}{r} \phi^{UV}_{r,r+2} - \frac{2}{r-1} \sqrt{\frac{r-2}{r}} (2 h^{UV}_{r,r})^{-1} \partial \bar \partial \phi^{UV}_{r,r} + \frac{2}{r(r-1)} \phi^{UV}_{r,r-2}\cr
 (2 h^{IR}_{r,r})^{-1} \partial \bar \partial \phi^{IR}_{r,r} &= \frac{2}{r+1} \sqrt{\frac{r+2}{r}} \phi^{UV}_{r,r+2}+\frac{r^2-5}{r^2-1} (2 h^{UV}_{r,r})^{-1} \partial \bar \partial \phi^{UV}_{r,r} - \frac{2}{r-1} \sqrt{\frac{r-2}{r}} \phi^{UV}_{r,r-2}
\end{align}

At higher order, one has to describe the mixing of Virasoro descendants of $\phi^{UV}_{r,r \pm n}$ for a finite range of values of $n$. They will flow to the same range of $\phi^{IR}_{r\pm n ,r}$ descendants, though the anomalous dimension information 
is not sufficient alone to disentangle different descendants at the same level of the same $\phi^{IR}_{t ,r}$.

\subsection{Expected properties of the RG domain wall} 
If we define the RG domain wall by a perturbative flow from the identity interface, 
it is clear that there are no relevant interface interactions which need to be turned on as we turn on $\phi_{1,3}$ in the bulk. 
Thus we are in a rather convenient setting where the RG domain wall should exist and be unique. 
 
 Because of the expected form of the RG flow of operators from the UV minimal model to the IR minimal model, 
we expect the RG domain wall to be a conformal boundary condition for $\CM_{p+1,p} \times \CM_{p,p-1}$ 
associated to a boundary state which lives inside the subspace
\begin{equation}
\sum_{t,r,s}^{ s-t \in 2\bZ} \left[ \phi^{IR}_{t,r}\right] \times \left[  \phi^{UV}_{r,s} \right] 
\end{equation}

The conformal perturbation theory analysis at large $k$ gives further constraints.  
If we take the leading order expression 
for some canonically normalized operator $O_i^{IR}$ in the IR theory 
as a linear combination of canonically normalized operators in the UV theory
\begin{equation}
O_i^{IR} = \sum_j b_{ij} O_j^{UV}
\end{equation}
we get a prediction for the disk one point functions of the RG boundary condition
\begin{equation}
\langle  \bar O_j^{IR} O_i^{UV}  | RG \rangle = b_{ij}
\end{equation}
at the leading order in the $k^{-1}$ expansion,
i.e.some coefficients in the boundary state 
\begin{equation}
| RG \rangle = \sum_{ij} b_{ij} |\bar O_j^{IR} O_i^{UV} \rangle
\end{equation}
Notice that the reflection trick involved in relating boundary conditions and interfaces 
requires $O_j^{IR} \to \bar O_j^{IR}$. Notice also that if we derived the $b_{ij}$ from the diagonalization of the perturbative anomalous dimension matrix,
we can only compute $b_{ij}$ up to 
multiplication on the left by a matrix which mixes the various descendants at the same level of 
a IR primary.

A priori, it is far from obvious that a valid boundary state exists, i.e. that
we can complete the $b_{ij}$ to functions $b_{ij}(k)$ which agree with $b_{ij}$ at the leading order 
and satisfy the usual modular constraints for a conformal boundary condition, giving a 
consistent spectrum of states for the theory on a segment.  

Indeed, it is not even completely obvious that the boundary state can satisfy 
the condition for conformal invariance
\begin{equation}
\left( L^{UV}_n + L^{IR}_n \right) | RG \rangle  = \left( \bar L^{UV}_{-n} + \bar L^{IR}_{-n} \right) | RG \rangle
\end{equation}
But it should be possible to use the Ward identities of three-point functions $C_{i,j,(1,3)}$ to show that at the leading order in $k^{-1}$ the RG
boundary state will be conformal, and will even be a permutation brane:
\begin{equation}
L^{UV}_n  | RG \rangle  = \bar L^{IR}_{-n} | RG \rangle \qquad \qquad L^{IR}_n  | RG \rangle  = \bar L^{UV}_{-n} | RG \rangle
\end{equation}

At the end of this section, we will propose an algebraic construction of $| RG \rangle$ and 
match it at the leading order in $k^{-1}$ to the explicit mixing coefficients computed above.  

\subsection{Topological interfaces and RG flow}
Topological interfaces offer a convenient way to recast the selection rules for the RG flow initiated by $\phi_{1,3}$.
A topological interface is an interface which commutes with the energy-momentum tensor, and is thus totally transmissive. A general discussion of topological interfaces in rational conformal field theories can be found in 
\cite{Frohlich:2006ch}.

Topological interfaces can be freely deformed in correlation functions. They have non-singular fusion among themselves and with other 
interfaces or boundary conditions. 
It is useful to remember that if an interface/boundary $\CI_a$ appears in the fusion of a topological interface $\CD$ and 
an interface/boundary $\CI_b$, the theory will admit a topological junction between $\CI_a$, $\CI_b$ and $\CD$. 

Topological interfaces which preserve a chiral algebra $\CA$ can end on twist fields, which transform in a pair of representations 
of $\CA$, $\bar \CA$ which may not be available for the usual bulk fields. In particular, in a RCFT $\CT_\CA$ defined 
by the diagonal modular invariant for the chiral algebra $\CA$, topological defects are given by Cardy's construction, 
and labeled by a representation $a$ of $\CA$. They can end on twist fields in representations $(a_1, \bar a_2)$ if $a$ 
appears in the fusion of $a_1$ and $a_2$. In particular they can end on purely chiral or purely anti-chiral twist fields 
in representation $a$ or $\bar a$. This fact will be crucial in our analysis.

Cardy's construction applied to topological interfaces (seen as ``permutation branes'' for $\CT_\CA \times \CT_\CA$) 
shows that the topological defect $\CD_a$ ``acts'' on a bulk field in representation 
$(a_1, \bar a_1)$ by multiplication by the familiar factor
\begin{equation}
d_{a,a_1} = \frac{S_{a, a_1}}{S_{1,a_1}}
\end{equation} 
Here one acts on the bulk field by surrounding it with a small loop of $\CD_a$. 
Form Verlinde's formula, it follows that the topological interfaces fuse accordingly to the usual fusion coefficients 
for $\CA$. 

There is a special class of topological interfaces which are called ``group-like'': 
a group-like interface has the property that fused with itself (more precisely its opposite) gives the identity interface. 
A nice property of group-like interfaces is that they can be swept across a bulk field $(a_1, \bar a_1)$, at the price of a $d_{a,a_1} $ multiplicative factor. 
Indeed, a group-like interface implements a symmetry of the CFT. 

A general topological interface $\CD$ swept across a bulk field $(a_1, \bar a_1)$ instead leaves behind a ``tail'':
it transforms the bulk field  to a sum over twist fields in the same representations, 
attached to a strand of any defect appearing in the fusion of $\CD$ with its opposite. 

\begin{figure}[htb]
\centering
\includegraphics[width=5in]{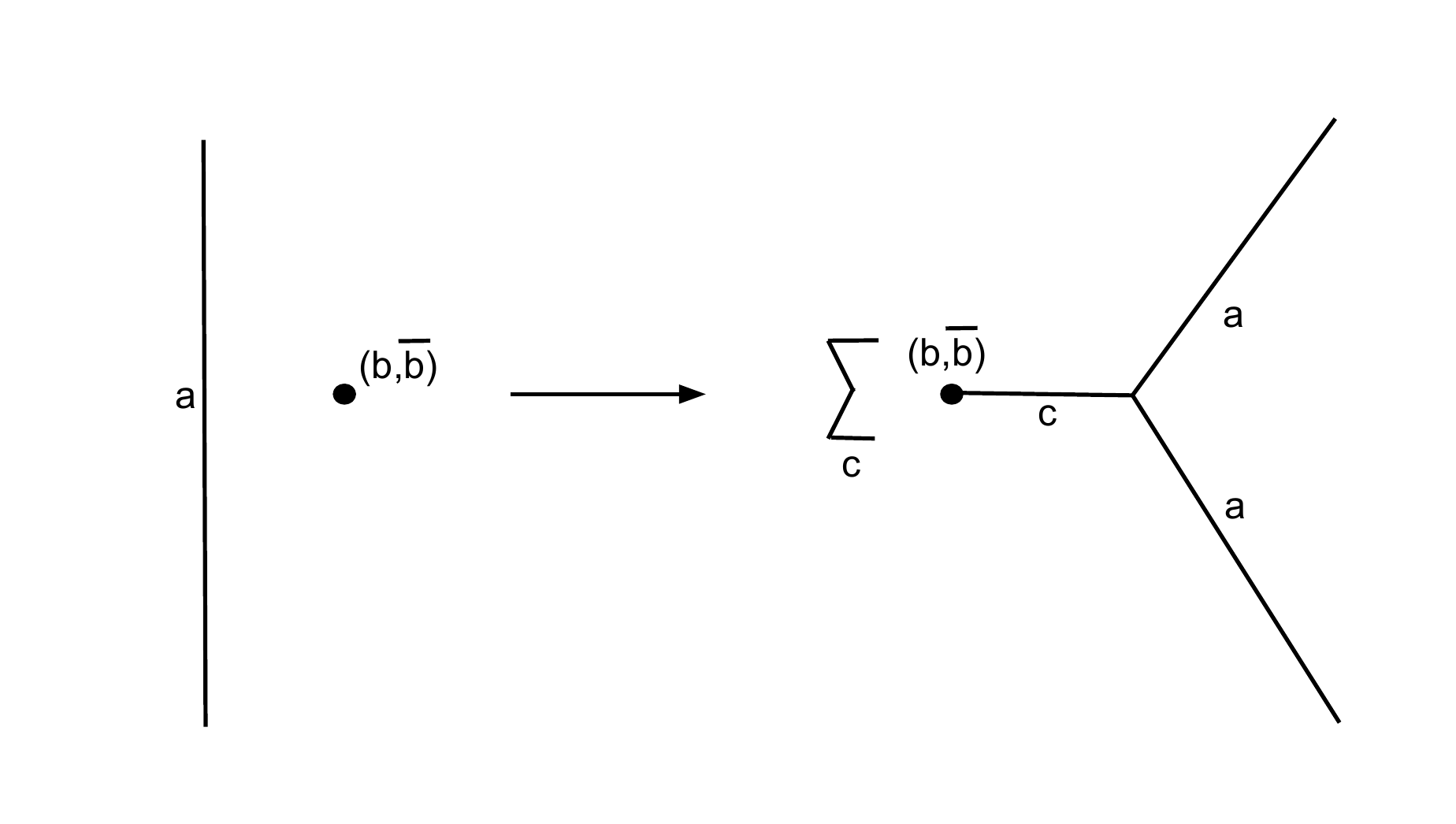}
\caption{A generic topological interface with label $a$ can be swept across a bulk field $(b, \bar b)$, but will transform it to a sum (with appropriate fusion coefficients) of twist fields attached to a segment of topological defect of label $c$. Notice that $c$ must be present in the fusion of $a$ with itself and 
in the fusion of $b$ and $\bar b$.}
\label{fig:four}
\end{figure}

We can use the Ising model to give some simple examples of topological interfaces. We refer to \cite{Frohlich} for more details and 
further references on the subject. Remember that the Ising model 
has three chiral primaries: the identity, $\sigma$ and $\epsilon$, of dimension $0$, $1/16$, $1/2$. 
The modular matrix is 
\begin{equation}S_{\mathrm{Ising}}= \left(
 \begin{array}{ccc}
 \frac{1}{2} & \frac{1}{\sqrt{2}} & \frac{1}{2} \\
 \frac{1}{\sqrt{2}} & 0 & -\frac{1}{\sqrt{2}} \\
 \frac{1}{2} & -\frac{1}{\sqrt{2}} & \frac{1}{2}
\end{array}
\right) \end{equation}

The topological defect $\CD_\epsilon$ labeled by the $h = 1/2$ primary in the Ising model is group-like: $\CD_\epsilon \CD_\epsilon =1$. 
It only acts non-trivially on the $\sigma(z,\bar z)$ primary field, multiplying it by $-1$: $\CD_\epsilon$ 
implements the spin-flip $Z_2$ symmetry of the Ising model. 

At the endpoint of a $\CD_\epsilon$ interface, one can place a chiral operator of dimension $1/2$, 
which is just the free fermion $\psi$ hidden in the Ising model.
The $\CD_\epsilon$ tail insures that the fermion is anti periodic around the spin field $\sigma(z,\bar z)$,
which is indeed one of the twist fields in the free-fermion description of the Ising model.
There is a second operator which can sit at the end of $\CD_\epsilon$: 
the disorder field $\mu(z,\bar z)$ of left and right conformal dimensions $1/16$,
i.e. the second twist field in the free fermion model. 

The second topological defect in the Ising model is $\CD_\sigma$. It is not group-like: indeed 
\begin{equation}
\CD_\sigma \CD_\sigma = 1 + \CD_\epsilon
\end{equation}
This topological defect is related to the high-low temperature duality in the Ising model: it changes the sign of $\epsilon$, 
and relates $\sigma$ and the disorder field $\mu$.  

After this simple example, we can look back at the $\CM_{p+1,p}$ minimal model. We want to understand what happens if we 
try to pull a $\CD_{r,1}$ defect across a $\phi_{1,s}$ operator as in figure \ref{fig:four}. The only representation which 
occurs both in the fusion of $(r,1)$ and itself and $(1,s)$ and itself is the identity representation. 
Thus the effect of sweeping the $\CD_{r,1}$ defect across a $\phi_{1,s}$ operator can at most be multiplication by a scalar factor. 
Because of rotation invariance, it is pretty clear that the factor should be $1$ or $-1$: sweeping the line defect  from the left to the right of the operator 
should be the same as sweeping from the right to the left. 

We can also compute the factor directly by comparing a small closed $\CD_{r,1}$ loop 
surrounding the operator and one which does not surround the operator, i.e. taking the ratio $d_{r,1;1,s}/d_{r,1;1,1}$. 
From the modular matrix 
\begin{equation}
S_{rs;r's'} = \sqrt{\frac{8}{p(p+1)}}(-1)^{(r+s)(r'+s')} \sin \frac{\pi r r'}{p} \sin \frac{\pi s s'}{p+1} 
\end{equation}
we compute 
\begin{equation}
d_{r,1;r's'} = (-1)^{(r-1)(r'+s')} \frac{\sin \frac{\pi r r'}{p}}{\sin \frac{\pi r'}{p} }
\end{equation}
and see that the ratio is $(-1)^{(r-1)(s-1)}$.  
Thus $\phi_{1,3}$ is completely transparent to the defects $\CD_{r,1}$: they require no renormalization in the perturbed theory, and should remain 
topological all the way to the IR minimal model. It is clear from the fusion rules that 
$\CD^{UV}_{r,1}$ coincides with $\CD^{IR}_{1,r}$ \cite{Fredenhagen:2009tn}.

\subsection{RG domain wall and topological defects}
As the relevant deformation $\phi$ is transparent 
to the topological defects  $\CD^{UV}_{r,1}$, the definition of the RG domain wall 
implies that the RG domain wall will be transparent to these topological defects.

More precisely, acting on the RG interface by $\CD^{UV}_{r,1}$ on one side should be the same 
as acting on the RG interface $\CD^{IR}_{1,r}$ on the other side.
Notice that $d^{UV}_{r,1;r's'}$ coincides with $d^{IR}_{1,r;t'r''}$ iff $r' = r''$ and 
$t'-s'$ is even. Requiring the RG domain wall to intertwine $\CD^{UV}_{r,1}$ and $\CD^{IR}_{1,r}$,
we recover the expected constraint that the boundary state $|RG \rangle$ 
 only contains total Virasoro Ishibashi states inside $ \left[ \phi^{IR}_{t,r}\right] \times \left[  \phi^{UV}_{r,s} \right] $  with even $t-s$ \cite{Fredenhagen:2009tn}. 

\begin{figure}[htb]
\centering
\includegraphics[width=6in]{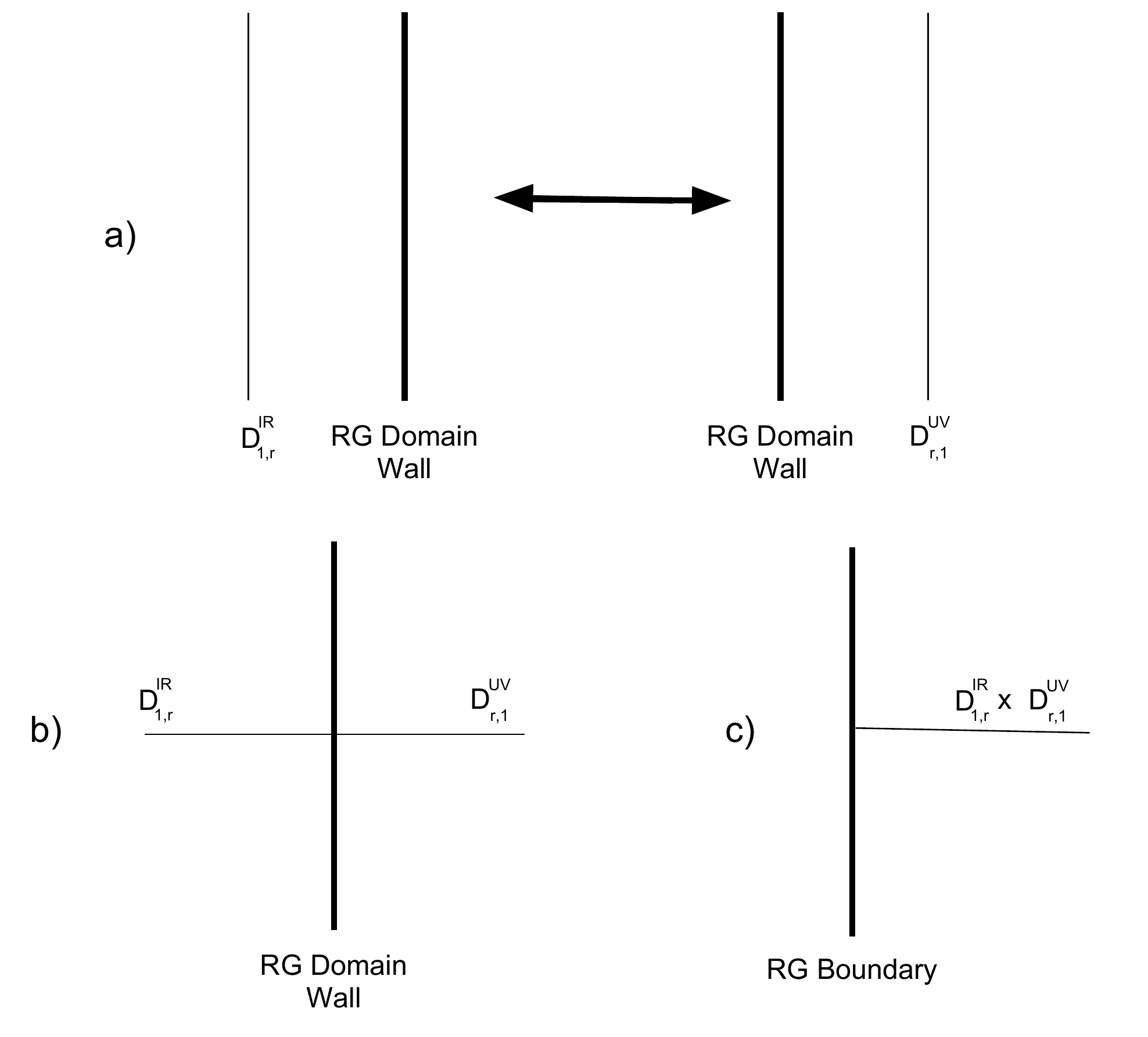}
\caption{a) The RG interface is expected to intertwine the topological defects $\CD^{UV}_{r,1}$ and $\CD^{IR}_{1,r}$. b) A $\CD^{UV}_{r,1}$ can continue topologically across a RG interface and become $\CD^{IR}_{1,r}$ on the other side.  c) Equivalently, the product defect 
$\CD^{IR}_{1,r}\CD^{UV}_{r,1}$ in the product theory $\CT_{UV} \times \CT_{IR}$ can end topologically on the RG boundary in a unique fashion.}
\label{fig:five}
\end{figure}

There is a subtle refinement of the intertwining property: 
it should be possible for a line defect $\CD^{UV}_{r,1}$ to cross topologically the RG interface, and become 
$\CD^{IR}_{1,r}$ on the other side.
This means that the combined defect  $\CD^{IR}_{1,r}\CD^{UV}_{r,1}$ in the product theory $\CT_{UV} \times \CT_{IR}$
can end topologically on the boundary defined by the RG interface. See Figure \ref{fig:five}

This will be a crucial observation. The OPE of a (anti)holomorphic operator with a boundary is always regular: 
the OPE coefficient can only be a function of the distance from the boundary, but it must also be (anti)holomorphic.
Thus it must be constant.
If we have a (anti)holomorphic twist field connected to the boundary by a topological defect, 
we can take the regular OPE of the twist field with the boundary: the topological defect disappears, and one is left with a 
standard boundary operator. 

We will do so momentarily with the RG boundary condition and the $\phi^{IR}_{1,r} \phi^{UV}_{r,1}$
(anti)holomorphic twist fields.  Before that, we can look at a simple example 
of topological defects ending on boundaries in the Ising model.

\begin{figure}[htb]
\centering
\includegraphics[width=6in]{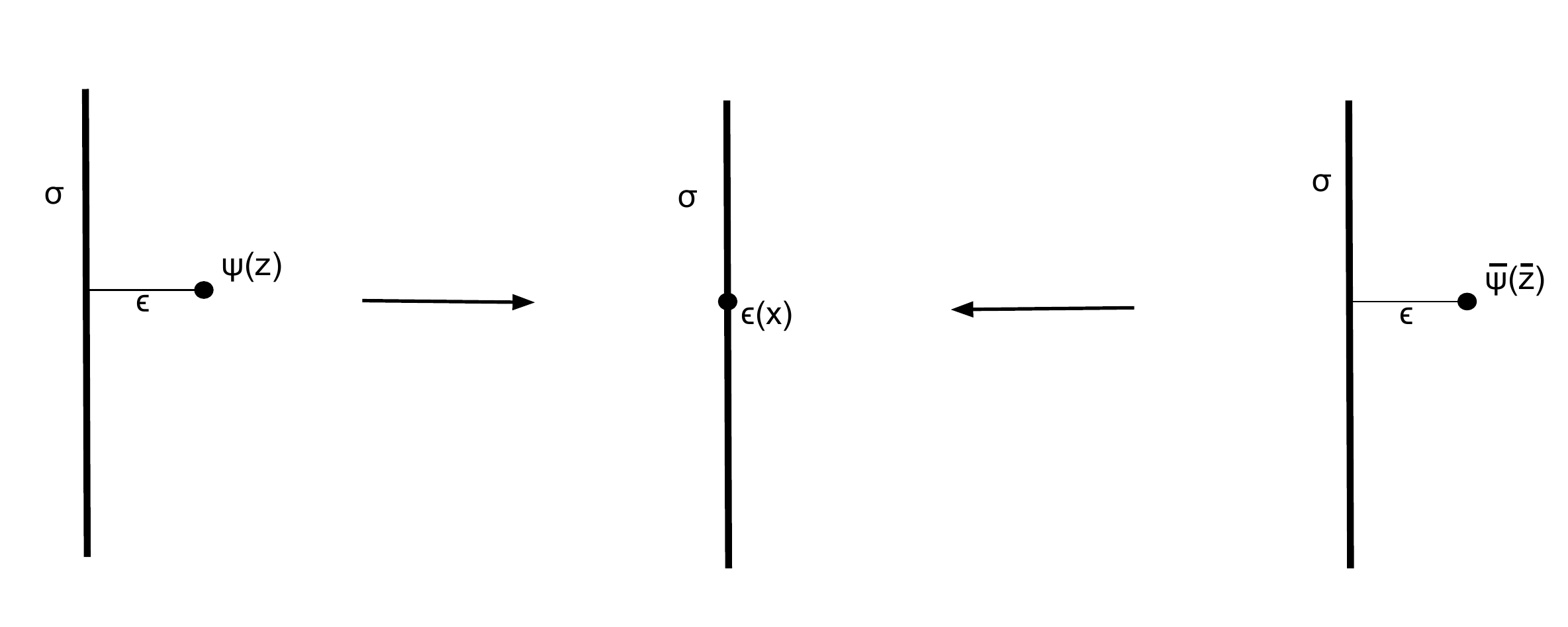}
\caption{As the topological defect labeled by $\epsilon$ can end topologically on a $\sigma$ boundary, we have a regular OPE of the holomorphic  twist field $\psi$ to a boundary field 
of dimension $1/2$, and a regular OPE of the anti-holomorphic  twist field $\bar \psi$ to the same boundary field. This gives a gluing condition for $\psi$ and $\bar \psi$ at the boundary.}
\label{fig:six}
\end{figure}

Because of the fusion rule $[\epsilon] \times [\sigma] = \sigma$, a $\CD_\epsilon$ defect can end 
topologically on the Cardy boundary condition labeled by $\sigma$. 
Thus we can consider the holomorphic fermion operator $\psi$ attached to a $\CD_\epsilon$ defect ending on the boundary,
and collide it with the boundary: the $\CD_\epsilon$ defect disappears, and one is left with a multiple 
of the unique dimension $1/2$ boundary operator which lives on the boundary labeled by $\sigma$. 
We can do the same with the anti-holomorphic fermion operator $\bar \psi$, and get a multiple of the same 
dimension $1/2$ boundary operator. This implies that the boundary condition labeled by $\sigma$ 
in the Ising model may be inherited from a simple boundary condition in the theory of free fermions,
which glue $\psi$ to a multiple of $\bar \psi$. See Figure \ref{fig:six}

\subsection{A hidden current algebra}
We will now argue that the holomorphic fields $\phi^{IR}_{1,r} \phi^{UV}_{r,1}$ with odd $r$ which live at the endpoints of 
$\CD^{UV}_{r,1}\CD^{IR}_{1,r}$ topological defects define a hidden extended current algebra $\CB$ in $\CT_{UV} \times \CT_{IR}$. Because the $\CD^{UV}_{r,1}\CD^{IR}_{1,r}$ defects can end topologically on the RG boundary condition,
we will be able to argue that the RG boundary condition is associated to a specific gluing map between the holomorphic
and anti-holomorphic copies of $\CB$. 

Take a closer look to the chiral fields $\phi^{IR}_{t,r} \phi^{UV}_{r,s}$ in the product theory. They have dimensions 
\begin{equation}
h^{IR}_{t,r} + h^{UV}_{r,s} = \frac{(r-s)^2}{4}+\frac{(t-r)^2}{4} + \frac{t^2-1}{4(k+1)} - \frac{s^2-1}{4(k+3)} 
\end{equation}
Notice that if we change $r$ by an even amount, the conformal dimension jumps by an integer. If we change it by an odd amount, the dimension 
jumps by an integer and a half. This suggests that the 
corresponding fields can be arranged into representations of some extended current algebra. 

The current algebra is readily identified: the fields $\phi^{IR}_{1,r} \phi^{UV}_{r,1}$ have integral dimension for odd $r$, half-integral for even $r$!
This set of fields is closed under fusion, and thus the product theory $\CT_{UV} \times \CT_{IR}$ has a hidden, extended chiral 
algebra realized by twist fields which live at the endpoints of $\CD^{UV}_{r,1}\CD^{IR}_{1,r}$ topological defects. 
We will refer to the algebra generated by integral currents, i.e. odd $r$, as $\CB$. 
It is also useful to consider the larger algebra $\tilde \CB$ which includes also the fermionic currents 
of even $r$. 
 
We can easily understand the origin of this hidden symmetry if we
remember that the unitary minimal models have a coset description 
\begin{equation}
{\cal M}_{k+3,k+2} = \frac{\hat {su}(2)_{k} \times \hat {su}(2)_1}{\hat {su}(2)_{k+1}}
\end{equation}

If we denote as $\phi^{(k)}_r$ the dimension $r$ primary of $\hat {su}(2)_{k}$, 
etc., the coset construction gives the decomposition of the tensor product of 
a $\hat {su}(2)_{k}$ and a $\hat {su}(2)_{1}$ representations as 
\begin{equation}
\left[ \phi^{(k)}_r \right] \times \left[ \phi^{(1)}_d \right] = \sum_s \left[\phi_{r,s} \right] \times \left[ \phi^{(k+1)}_s \right]
\end{equation}
where $s+r+d$ must be odd. Thus the Virasoro primary $\phi_{r,s}$ coincides with the coset field labeled by 
$[r,d;s]$, where $d=1$ if $r-s$ is even, $d=2$ otherwise.

It is also useful to remember that we can write the minimal model modular matrix for the minimal model 
in terms of the modular matrix for the WZW model 
\begin{equation}
S^{(k)}_{n,m} = \sqrt{\frac{2}{k+2}} \sin \frac{\pi n m}{k+2}
\end{equation}
 as 
\begin{equation}
S_{[r,d;s],[r',d';s']} = 2 S^{(k)}_{r,r'} S^{(1)}_{d,d'} S^{(k+1)}_{s,s'}
\end{equation}
where the factor of $2$ accounts for the selection rules and identifications for the coset fields. 

Now we can look at the coset description of the product theory $\CT^{IR} \times \CT^{UV}$: 
\begin{equation}
{\cal M}_{k+2,k+1} \times {\cal M}_{k+3,k+2} =  \frac{\hat {su}(2)_{k-1} \times \hat {su}(2)_1}{\hat {su}(2)_{k}} \times \frac{\hat {su}(2)_{k} \times \hat {su}(2)_1}{\hat {su}(2)_{k+1}}
\end{equation}
and compare it with the theory $\CT_\CB$ defined by the diagonal modular invariant for the coset 
\begin{equation}
\CB = \frac{\hat {su}(2)_{k-1} \times \hat {su}(2)_1 \times \hat {su}(2)_1}{\hat {su}(2)_{k+1}}
\end{equation}

In the first coset, we decompose
\begin{equation}
\left[ \phi^{(k-1)}_t \right] \times \left[ \phi^{(1)}_d \right] \times \left[ \phi^{(1)}_{\tilde d} \right] = \sum_{r,s} \left[\phi^{IR}_{t,r} \right] \times \left[\phi^{UV}_{r,s} \right] \times \left[ \phi^{(k+1)}_s \right]
\end{equation}
In the latter coset we decompose
\begin{equation}
\left[ \phi^{(k-1)}_t \right] \times \left[ \phi^{(1)}_d \right] \times \left[ \phi^{(1)}_{\tilde d} \right] = \sum_{s} \left[\phi^{\CB}_{[t,d,\tilde d;s]} \right] \times \left[ \phi^{(k+1)}_s \right]
\end{equation}
Thus the $\CB$ representations labeled by $[t,d,\tilde d,s]$
is simply the direct sum of representations of ${\cal M}_{k+2,k+1} \times {\cal M}_{k+3,k+2}$ with labels 
$[t,d,r]$ and $[r, \tilde d,s]$. 

For a given pair $t$,$s$, we have two choices of values for the pair $d$, $\tilde d$,
corresponding to the sum over even and odd values of $r$ respectively.  
These pairs of $\CB$ representations form a single $\tilde \CB$ representation labeled by $[t,s]$. 
The primaries with even $t-s$ are in NS sectors for the fermionic currents of $\tilde \CB$,
while the primaries with odd $t-s$ are in Ramond sectors.

The theory $\CT_\CB$ which enjoys the full current algebra symmetry is not the same as the product theory 
$\CT_{UV} \times \CT_{IR}$. Rather, it should be thought of as a different modular invariant 
for the chiral algebra of the product theory. For notational convenience, we will often denote the product theory as 
\begin{equation}
\CT_\CA = {\cal M}_{k+2,k+1} \times {\cal M}_{k+3,k+2}
\end{equation} 
and the corresponding chiral algebra as $\CA$. 

The identity representation of $\CB$
\begin{equation}
\phi_{[1,1,1,1]} = \sum_{r \in 2 \bZ+1} \left[ \phi^{IR}_{1,r}\right] \times \left[ \phi^{UV}_{r,1} \right] 
\end{equation}
contains three fields of conformal dimension $2$: the energy-momentum tensors 
$T^{UV}$ and $T^{IR}$, but also the interesting current $\phi^{IR}_{1,3} \phi^{UV}_{3,1}$.

The chiral algebra $\CB$ has an obvious $Z_2$ symmetry which exchanges the 
two factors $\hat {su}(2)_1 \times \hat {su}(2)_1$ in the numerator of the coset. 
The $Z_2$ symmetry exchanges two inequivalent embeddings of $\CA$ into $\CB$. 
In particular, it will mix among themselves $T^{UV}$, $T^{IR}$ and $\phi^{IR}_{1,3} \phi^{UV}_{3,1}$!
This means that the $Z_2$ symmetry acts in a very non-trivial way on each $\CB$ 
representation, mixing in an intricate way different $\CA$ representations. 

It is also useful to look at the current $\psi = \phi^{IR}_{1,2} \phi^{UV}_{2,1}$, of dimension $1/2$, which sits in the identity representation for $\tilde B$. 
This is actually a free-fermion: the OPE $\psi \psi$ only contains the identity and fields of dimension $2$ or higher.  
Indeed we can give another, powerful, description of the $\tilde \CB$ current algebra:
it is just the direct product of the free fermion algebra generated by $\psi$ and 
a super Virasoro algebra! 

This observation, and many of the explicit formulae we derive from it, are discussed in detail in the reference \cite{Crnkovic:1989ug}. It follows from the coset construction
\begin{equation}
\CS \CM_{k+1,k+3} \times \CM_{4,3} = \frac{\hat {su}(2)_{k-1} \times \hat {su}(2)_2}{\hat {su}(2)_{k+1}} \times \frac{\hat {su}(2)_1 \times \hat {su}(2)_1}{\hat {su}(2)_{2}}
\end{equation}
The second factor is just the Ising model associated to the fermionic current $\psi = \phi^{IR}_{1,2} \phi^{UV}_{2,1}$.
 The first factor is a supersymmetric minimal model. 
 This coset makes it obvious that the current algebra $\CB$ consists of the currents of integral dimension built from the super-Virasoro algebra and $\psi$, and that the $Z_2$ automorphism acts simply as $\psi \to -\psi$.  

The $\CB$ representations labeled by $[t,d,\tilde d,s]$ decompose into Ising model representations with label $[d, \tilde d, d']$ and supersymmetric minimal model representations 
$[t, d',s]$. Here $d' = 1,2,3$. Parsing through definitions, we see that the $\CB$ representations either combine NS representations of the supersymmetric minimal model and the $1$, $\epsilon$ 
fields of the Ising model, or Ramond representations and the spin field $\sigma$. Indeed, $\CT_\CB$ can be seen as the orbifold of $\CS \CM_{k+1,k+3} \times \CM_{4,3}$ 
by the $Z'_2$ symmetry which reflects Ramond operators and the spin field $\sigma$. 

The three dimension $2$ currents in $\CB$ are simply a linear combination of the energy-momentum tensors $T$ and $T_\psi$ of $\CS \CM_{k+1,k+3} \times \CM_{4,3}$ and of the combination 
$\psi G$ of free fermion current and superconformal generator. As we know fully the OPE of these fields, we can solve for the expressions of $T_{UV}$ and $T_{IR}$: 
\begin{align}
T^{IR} &= \frac{k+3}{2 k+4} T + \frac{ \sqrt{(k+1)(k+3)} }{2k+4}G \psi + 
  \frac{k-1}{2 k+4}  T_\psi \cr
T^{UV} &= \frac{k+1}{2 k+4} T -\frac{ \sqrt{(k+1)(k+3)} }{2k+4}G \psi +
   \frac{k+5}{2 k+4} T_\psi
\end{align}
The third current is
\begin{equation}
\phi^{IR}_{1,3} \phi^{UV}_{3,1}= (k+1) (k+3) T - 3\sqrt{(k+1)(k+3)}G  \psi - 3 (k-1) (k+5) T_\psi
\end{equation}
We can confirm that indeed this field has conformal dimension $\frac{k}{k+2}$ under $T^{IR}$, and hence $\frac{k+4}{k+2}$ under $T^{UV}$. 
As written, it is not canonically normalized. It should be divided by the square root of 
$3 (k-1) k ( k+4) (k+5)$ to be normalized to $1$.

In these and later computations, we let the super-Virasoro currents and $\psi$ commute with each other. 
It may be more natural to let $G$ and $\psi$ anti-commute, as they are both fermionic. 
Then one would need to put a factor of $i$ in front of bilinears such as $G \psi$
to make the formulae work. 
 
The $Z_2$ symmetry of $\CB$ acts as $\psi \to - \psi$. It maps $T^{IR}$ and $T^{UV}$ respectively to  
\begin{align}
\tilde T^{IR} &= \frac{3}{(k+2) (k+4)}T^{IR}+\frac{(k-1) (k+3)}{k (k+2)}T^{UV} + \frac{1}{k (k+2)
   (k+4)}\phi^{IR}_{1,3} \phi^{UV}_{3,1} \cr
\tilde T^{UV} &=  \frac{(k+1) (k+5)}{(k+2) (k+4)}T^{IR}+\frac{3}{k (k+2)}T^{UV} -\frac{1}{k (k+2)
   (k+4)}\phi^{IR}_{1,3} \phi^{UV}_{3,1}
\end{align}

Given sufficient patience, or some Mathematica code to deal with the free fermion and super-Virasoro algebra, one can
expand any descendant of operators of the form $\phi^{IR}_{t,r} \phi^{UV}_{r,s}$ in the product theory in terms of descendants 
of super-Virasoro primaries and Ising model fields. The algorithm is conceptually straightforward: 
take general linear combinations of the operators at the same level in the same $\CB$ representation, and 
act with the explicit $T^{IR}$ and $T^{UV}$ to identify the primary fields $\phi^{IR}_{t,r} \phi^{UV}_{r,s}$.
This allows one to compute explicitly the $Z_2$ action on descendants of $\phi^{IR}_{t,r} \phi^{UV}_{r,s}$.

\subsection{The extended current algebra of the RG domain wall}
The total holomorphic and anti-holomorphic energy-momentum tensors brought to the boundary 
give the same operator, the boundary energy-momentum tensor. But the individual energy-momentum tensors 
of $\CT^{UV}$ and $\CT^{IR}$ will not match with their anti-holomorphic versions when brought to the RG boundary condition. 
Rather, we have two copies of $\CA$ at the RG boundary, 
which intersect along the common energy-momentum tensor. In particular, there must be at least three dimension $2$ currents.

There is a striking consequence of the fact that $\CD^{IR}_{1,r}\CD^{UV}_{r,1}$ can end on the boundary condition defined by the RG domain wall.
The holomorphic currents $\phi^{IR}_{1,r} \phi^{UV}_{r,1}$ can sit at the end of a $\CD^{IR}_{1,r}\CD^{UV}_{r,1}$ defect 
attached to the boundary, and brought to the boundary to complete $\CA$ to a full copy of the $\CB$ algebra. A second copy of $\CB$ arises from the anti-holomorphic currents.  
In principle, it is possible that these two copies of $\CB$ are distinct and intersect only on the overall Virasoro algebra. 
But it is more economical to conjecture that they actually coincide. 
The two copies of $\CA$ at the RG boundary condition may simply coincide with the two copies of $\CA$ inside $\CB$,
related by the $Z_2$ symmetry. 

More generally, we can place a holomorphic current $\phi^{IR}_{t,r} \phi^{UV}_{r,s}$ at the tip of a $\CD^{IR}_{t,1} \CD^{UV}_{1,s}$ 
line defect coming from infinity, and connect it by $\CD^{UV}_{r,1}\CD^{IR}_{1,r}$ to the boundary. 
By OPE with the boundary, we get a copy of each $\CB$ representation from the holomorphic currents attached to $\CD^{IR}_{t,1} \CD^{UV}_{1,s}$, and another copy from the anti-holomorphic 
currents. Again, the most economic choice is to identify these two copies under the $Z_2$ automorphism of $\CB$.

Notice that at the perturbative level, this is a reasonable assumption. The interface operators for the RG domain wall arise from the perturbative renormalization of the 
local operators at the trivial interface for $\CT^{UV}$, which are just the bulk fields. It is already rather surprising that enough of these operators would be renormalized 
to interface fields with exactly integral dimension to fill in one copy of $\CB$. It would take even more surprising cancellations to give two copies of $\CB$. 

At this point, we have the first non-trivial check: the $Z_2$ automorphism which relates the four energy-momentum tensors and the two $\phi^{IR}_{1,3} \phi^{UV}_{3,1}$
currents must be perturbatively close to the trivial gluing condition of the trivial interface in $\CT_{UV}$. 
Indeed, inspection shows that at the leading order in $k^{-1}$, $\tilde T^{IR} \sim T^{UV}$ and $\tilde T^{UV} \sim T^{IR}$!
Remember that a relation between $\bar T^{IR}$ and $T^{UV}$ at the RG boundary condition corresponds to a
relation between $T^{IR}$ and $T^{UV}$ at the RG interface because of the complex conjugation involved in 
relating interfaces and boundary conditions. 

If the RG boundary condition was a brane in the theory $\CT_\CB$, the gluing condition for the holomorphic and antiholomorphic copies of $\CB$ 
would imply that it is a rational brane. On the other hand, the RG boundary condition is a brane in the product theory $\CT_\CA$.
In order to make full use of the $\CB$ symmetry, we should relate the RG boundary condition to a brane in the theory $\CT_\CB$. 

In the next section, we will simply pick the simplest $Z_2$-twisted boundary condition in the theory $\CT_\CB$, and map it to 
the product theory $\CT_\CA$ by acting with the simplest $\CA$-topological interface between $\CT_\CB$ and $\CT_\CA$.
We will show that the resulting boundary condition has all the properties which we associated to the RG boundary condition: 
\begin{itemize}
\item it is a general, non-rational conformal boundary condition 
\item the boundary state involves only total Virasoro Ishibashi states inside $ \left[ \phi^{IR}_{t,r}\right] \times \left[  \phi^{UV}_{r,s} \right] $ with even $t-s$ 
\item the $\CD^{IR}_{1,r}\CD^{UV}_{r,1}$ topological defects can end on it in a natural way 
\item It glues the holomorphic and antiholomorphic copies of $\CB$ by the $Z_2$ automorphism
\end{itemize}
Thus we will take this boundary condition as our candidate RG domain wall. 

\subsection{Topological interfaces between theories}
There are some general facts which are true whenever one has a 2d RCFT $\CT_{\CB}$ which is defined by the diagonal modular invariant 
for a chiral algebra $\CB$, and is also a RCFT under a subalgebra $\CA \subset \CB$. 

Concretely, this means that the characters $\chi_\mu$ for representations $\mu$ of $\CB$ can be decomposed as a finite sum of characters 
$\chi_a$ for representations $a$ of $\CA$
\begin{equation}
\chi_\mu = \sum_a n^a_\mu \chi_a
\end{equation}
There is a useful relation between the $S$ modular matrices of the two chiral algebras:
\begin{equation}
n^a_\mu S_a^b = S_\mu^\nu n^b_\nu
\end{equation}
For simplicity we will assume that all representations are self-conjugate, and that $n^a_\mu$ takes values $0$ or $1$. 
This is surely true in our current example.

This setup admits several interesting topological defects. There is of course the Cardy basis of $\CA$-topological defects
for $\CT_{\CA}$:
\begin{equation}
\CD_a = \sum_b \frac{S^b_a}{S^b_1} \|\CA,b\|
\end{equation}
where we denote as $\|\CA,b\|$ the $\CA$-Ishibashi state which projects to the representation $b$ (and $\bar b$)
and commutes with $\CA$ (and $\bar \CA$). 
Similarly, there is also a Cardy basis of $\CB$-topological defects
for $\CT_{\CB}$:
\begin{equation}
\tilde \CD_\mu = \sum_\mu \frac{S^\nu_\mu}{S^\nu_1} \|\CB,\nu\| 
\end{equation}

We are interested in $\CA$-topological interfaces between $\CT_{\CA}$ and $\CT_{\CB}$.
It is known that the number of such interfaces is the same as the trace 
\begin{equation}
\sum_{\mu,a} n^a_\mu n^a_\mu
\end{equation}
i.e. there are as many interfaces as pairs $\mu$,$a$ allowed by the branching rules. 
One can write a general interface as
\begin{equation}
\CI_x = \sum_{\mu,a} \frac{S^{\mu,a}_x}{S^a_1} \|\CA,\mu,a\| 
\end{equation}
where  $\|\CA,\mu,a\|$ is the $\CA$-Ishibashi state which pairs up the representation $a$ in $\CT_{\CA}$ and 
the copy of $a$ inside $\mu$ in $\CT_{\CB}$. We will denote as $\tilde \CI_x$ the same interface running in the opposite direction.

The coefficients $S^{\mu,a}_x$ are constrained by the requirement that $ \tilde \CI_y \CI_x$ should be a direct sum of topological defects in 
$\CT_{\CA}$:
\begin{equation}
\tilde \CI_y \CI_x = \sum_{\mu,a}  \frac{S^{\mu,a}_x}{S^a_1} \frac{S^{\mu,a}_y}{S^a_1} \|\CA,a\| = \sum_b N_{x,y}^b \CD_b
\end{equation}
i.e.
\begin{equation}
N_{x,y}^b = \sum_{\mu,a} \frac{S^{\mu,a}_x S^{\mu,a}_y S_a^b}{ S^a_1} 
\end{equation}
and that the defects should be elementary, i.e. $N_{x,y}^1 = \delta_{x,y}$.

It is possible, but intricate, to solve these constraints. 
But there is a special $\CA$-topological interface which can be written down immediately:
\begin{equation}
\CI_1 = \sum_{\mu,a} \sqrt{\frac{S^\mu_1}{S^a_1}} \|\CA,\mu,a\| 
\end{equation}
Indeed, 
\begin{equation}
\tilde \CI_1 \CI_1 = \sum_{\mu,a} n_\mu^a \frac{S^\mu_1}{S^a_1} \|\CA,a\|  = \sum_b n_1^b \CD_b 
\end{equation}

The latter equality implies that the $\CD_a$ topological defects can end topologically 
on $\CI_1$. In particular, the holomorphic currents in $\CB$ can be 
pulled through $\CI_1$ from the $\CT_\CB$ side to the $\CT_\CA$ side, 
attached to a $\CD_a$ tether. 

Armed with an explicit topological interface, we can map boundary conditions in $\CT_\CB$ to 
(superpositions of) boundary conditions in $\CT_\CA$ and viceversa. 
This is especially interesting if the algebra $\CB$ has an automorphism $g$ which does {\it not} fix $\CA$. 
Then a rational brane $\tilde B^g$ for $\CT_\CB$ which identifies the holomorphic and 
anti-holomorphic copies of $\CB$ with the gluing map $g$ for $\CB$ will be mapped under 
the action of $\tilde \CI_1$ to a brane $B^g = \tilde \CI_1 \tilde B^g$ for $\CT_\CA$ which only preserves the intersection $\CA \cap g(\CA)$,
which can consist, say, of the stress tensor only.  

\begin{figure}[htb]
\centering
\includegraphics[width=6in]{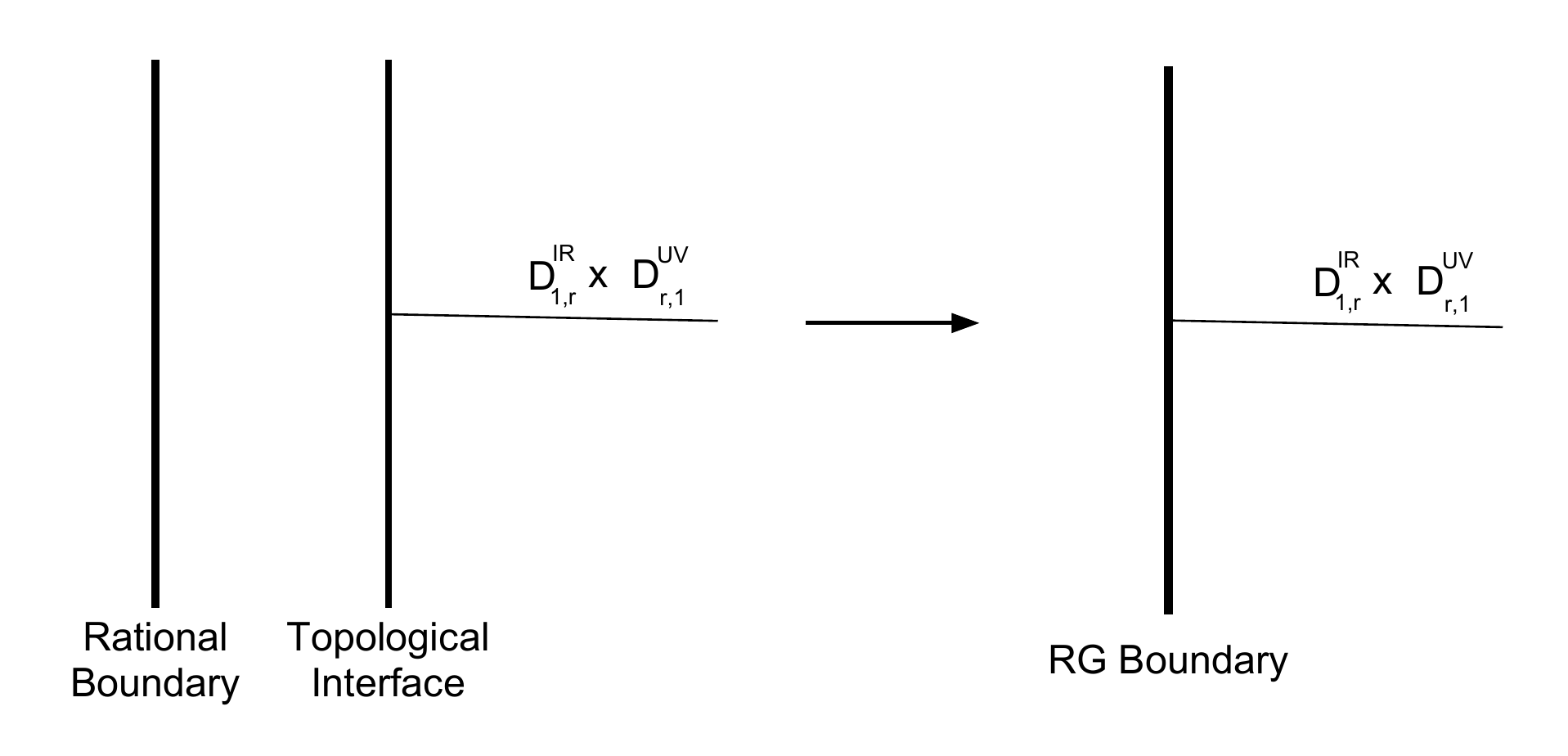}
\caption{The topological defects $\CD^{IR}_{1,r}\CD^{UV}_{r,1}$ can end on the topological interface $\CI_1$. Thus they will be able to end topologically on the candidate RG boundary.  }
\label{fig:seven}
\end{figure}

Thus we get a rather explicit definition
of a class of conformal, but not rational, boundary conditions for $\CT_\CA$. 
Notice that $\CD_a$ topological defects for $a$ in the identity representation of $\CB$ can end naturally on $B^g$:
they just end on $\tilde \CI_1$. Furthermore, the holomorphic and anti-holomorphic 
$\CB$ currents in the theory $\CT_\CA$, attached to a defect ending on $\tilde \CI_1$ can be brought through $\CI_1$ to become 
standard currents in the $\CT_\CB$ theory, and then related by the $g$ automorphism when brought to the boundary 
$B^g$. See Figure \ref{fig:seven}

Specializing to our setup, $\mu$ runs over possible $[t, d, \tilde d, s]$ and the $a$ in such $\mu$ 
run over $[t,d,r] \times [r,\tilde d, s]$, with $d$ and $\tilde d$ fixed by the parity of $t+r$ and $r+s$. 
Explicitly, 
\begin{equation}
\CI_1 = \sum_{t,r,s} \frac{1}{S^{(k)}_{1,r}} \|\CA,t,r,s\| 
\end{equation}

The $\CI_1$ image of any $\CT_\CB$ brane would only involve descendants of 
combinations $\phi^{IR}_{t,r} \phi^{UV}_{r,s}$. A $Z_2$ twisted brane in $\CT_\CB$
only involves Ishibashi states built on primaries with $d =\tilde d$ labels. 
If $t-s$ is odd, the labels $d$ and $\tilde d$ must differ.
Thus the $\CI_1$ images of $Z_2$-twisted $\CT_\CB$ branes satisfy the same selection rules 
as expected from the RG boundary condition: $t-s$ should be even.  

The simplest $Z_2$-twisted brane in $\CT_\CB$ takes the  form
\begin{equation}
|\tilde B \rangle = \sum_{t,s}^{s-t \in 2 Z} \sqrt{S^{(k-1)}_{1,t} S^{(k+1)}_{1,s}} | t, s; \tilde \CB, Z_2 \rrangle 
\end{equation}
where the $Z_2$-twisted $\tilde \CB$ Ishibashi state can be rewritten as the sum of two $Z_2$-twisted $\CB$ Ishibashi states, or the product 
of a super-Virasoro Ishibashi state and an Ising model Ishibashi state:
\begin{align} | t, s; \tilde \CB, Z_2 \rrangle &=| t,1,1, s;\CB, Z_2 \rrangle+ | t,2,2, s;\CB, Z_2 \rrangle \cr
&= \big(|t,1,s \rrangle + |t,3,s  \rrangle\big) \times \big( |1 \rrangle- | \epsilon \rrangle \big) 
\end{align} 
Indeed, the brane $|\tilde B \rangle$ is the $Z_2'$ orbifold image of the brane labeled by the identity in 
the super-Virasoro minimal model tensored with the brane labeled by $\sigma$ in the Ising model.  

Acting with $\CI_1$, we get our candidate
\begin{equation}
|RG \rangle = \sum_{t,s,r}^{s-t \in 2 Z} \frac{\sqrt{S^{(k-1)}_{1,t} S^{(k+1)}_{1,s}}}{S^{(k)}_{1,r}} \|\CA,t,r,s| | t, s; \tilde \CB, Z_2 \rrangle 
\end{equation}

We can give an explicit recipe to compute the one point function
\begin{equation}
\langle O^{UV} \bar O^{UV} O^{IR} \bar O^{IR} |RG\rangle
\end{equation}
of any product of descendants $O^{UV} \bar O^{UV}$ and $O^{IR} \bar O^{IR}$ of 
some $\phi^{IR}_{t,r} \phi^{UV}_{r,s}$:
\begin{itemize}
\item Translate $O^{UV} O^{IR}$ as a certain $\tilde \CB$ descendant of the $\tilde \CB$ primary $\phi_{t,s}$. 
\item Translate $\bar O^{UV} \bar O^{IR}$ as a certain $\tilde \CB$ descendant of the $\tilde \CB$ primary $\phi_{t,s}$.  
\item Using the $\CB$ current algebra, compute the overlap between $\bar O^{UV} \bar O^{IR}$ and the $Z_2$ image of $O^{UV} O^{IR}$
\item Multiply the overlap by $\frac{\sqrt{S^{(k-1)}_{1,t} S^{(k+1)}_{1,s}}}{S^{(k)}_{1,r}}$
\end{itemize}

\subsection{Explicit calculations} 
This description of $|RG \rangle$ is particularly useful because the $\tilde \CB$ current algebra  
is given explicitly by the product of a free fermion and super-Virasoro. 
 For a concrete calculation of a non-zero disk one-point function of an operator $O = O^{UV} \times O^{IR}$ 
 in the $[t,r,s]$ sector of $\CT_{UV} \times \CT_{IR}$, 
we need to represent $O$ explicitly in terms of NS fields in the supersymmetric minimal model 
and $\epsilon$ in the Ising model, by using the explicit expressions of $T^{UV}$ and $T^{IR}$ given above. 

Let's apply this procedure for some explicit comparisons with perturbative expectations.
It is easiest to start with fields in $\CT^{UV}$ which are not mixed at the leading order in perturbation theory:
the $\phi^{UV}_{r,r}$ fields with finite $r$, which are expected to flow to $\phi^{IR}_{r,r}$. Both fields have dimension close to zero.
The combined field $\phi^{IR}_{r,r} \phi^{UV}_{r,r}$ is the bottom component of 
the $[r,1,1,r]$ representation of $\CB$.
That bottom component coincides with the 
NS field labeled by $[r,1,r]$ in the supersymmetric minimal model.\footnote{As a check, notice that 
\begin{equation} h^{UV}_{r,r} = \frac{r^2-1}{4(k+2)(k+3)} \qquad  h^{IR}_{r,r} = \frac{r^2-1}{4(k+2)(k+1)}
\end{equation}
agree with the expressions of $T^{UV}$ and $T^{IR}$: they coincide with the conformal dimension in the supersymmetric minimal model 
multiplied by the coefficient of $T$ in $T^{UV}$ and $T^{IR}$.} The $Z_2$ symmetry acts trivially on this field.

Thus the one point function is simply 
\begin{equation}
\langle \phi^{IR}_{r,r} \phi^{UV}_{r,r} | RG \rangle =  \frac{\sqrt{S^{(k-1)}_{1,r} S^{(k+1)}_{1,r}}}{S^{(k)}_{1,r}} 
\end{equation}
This is $1$ up to terms of order $k^{-2}$, as desired. 

For a less boring example, we can look at the pair of fields $\phi^{UV}_{r,r\pm 1}$, which are known to mix in degenerate perturbation theory 
to give $\phi^{IR}_{r\pm 1,r}$. Now, $\phi^{IR}_{r- 1,r} \phi^{UV}_{r,r+ 1}$ and $\phi^{IR}_{r+ 1,r} \phi^{UV}_{r,r-1}$ are the bottom components 
respectively of the $[r+1,2,2,r-1]$ and $[r-1,2,2,r+1]$ representations of $\CB$. In turns, these bottom components coincides with the 
NS field labeled by $[r+1,3,r-1]$ and $[r-1,3,r+1]$ in the supersymmetric minimal model. Thus the $Z_2$ action is trivial. 

On the other hand, $\phi^{IR}_{r- 1,r} \phi^{UV}_{r,r-1}$ and $\phi^{IR}_{r+ 1,r} \phi^{UV}_{r,r+1}$ are 
linear combinations of the two level $1/2$ $\tilde \CB$ descendants of the NS fields labeled  
$[r-1,1,r-1]$ and $[r+1,1,r+1]$ respectively.
Thus we have an interesting diagonalization problem at hand: find the linear combination of primaries 
 $\epsilon\phi_{s,1,s} $ and $\phi_{s,3,s} \equiv G_{-1/2} \phi_{s,1,s}$ which is a primary field for $T^{UV}$. 
 
 The total conformal dimension is 
 \begin{equation}
h_{s,s} + \frac{1}{2}=  \frac{1}{2} + \frac{s^2-1}{2(k+3)(k+1)}
 \end{equation}
 We can compute
 \begin{equation}
 L_0^{UV} \epsilon \phi_{s,1,s}  =\left( \frac{s^2-1}{4(k+3)(k+2)}+ \frac{k+5}{4( k+2)}\right) \epsilon \phi_{s,1,s}   -\frac{ \sqrt{(k+1)(k+3)} }{2k+4}G_{-1/2} \phi_{s,1,s}
 \end{equation}
and 
 \begin{equation}
 L_0^{UV} G_{-1/2} \phi_{s,1,s} =\left( \frac{s^2-1}{4(k+3)(k+2)}+ \frac{k+1}{4( k+2)}\right) G_{-1/2} \phi_{s,1,s}  -  \frac{s^2-1}{2(k+2)\sqrt{(k+1)(k+3)}}  \epsilon \phi_{s,1,s} 
 \end{equation}

We find, properly normalized, 
 \begin{equation}
\phi^{IR}_{s,s-1} \phi^{UV}_{s-1,s} = \frac{(s-1)}{\sqrt{2 s (s-1)}}\epsilon \phi_{s,1,s} +  \frac{\sqrt{(k+1)(k+3)}}{\sqrt{2 s (s-1)}} G_{-1/2} \phi_{s,1,s}
 \end{equation}
 and 
 \begin{equation}
\phi^{IR}_{s,s+1} \phi^{UV}_{s+1,s} = \frac{(s+1)}{\sqrt{2 s (s+1)}}\epsilon \phi_{s,1,s} - \frac{\sqrt{(k+1)(k+3)}}{\sqrt{2 s (s+1)}} G_{-1/2} \phi_{s,1,s}
 \end{equation}
The overlap between $\phi^{IR}_{s,s-1} \phi^{UV}_{s-1,s}$ and its $Z_2$ image is simply $s^{-1}$, and  the overlap between $\phi^{IR}_{s,s+1} \phi^{UV}_{s+1,s}$ and its $Z_2$ image  is simply $-s^{-1}$.

Thus we have the following one-point functions: 
\begin{align}
\langle \phi^{IR}_{r-1,r} \phi^{UV}_{r,r+1} | RG \rangle &=  \frac{\sqrt{S^{(k-1)}_{1,r-1} S^{(k+1)}_{1,r+1}}}{S^{(k)}_{1,r}} \cr
\langle \phi^{IR}_{r+1,r} \phi^{UV}_{r,r-1} | RG \rangle &=  \frac{\sqrt{S^{(k-1)}_{1,r+1} S^{(k+1)}_{1,r-1}}}{S^{(k)}_{1,r}} \cr
\langle \phi^{IR}_{r-1,r} \phi^{UV}_{r,r-1} | RG \rangle &= -\frac{1}{r-1} \frac{\sqrt{S^{(k-1)}_{1,r-1} S^{(k+1)}_{1,r-1}}}{S^{(k)}_{1,r}} \cr
\langle \phi^{IR}_{r+1,r} \phi^{UV}_{r,r+1} | RG \rangle &=  \frac{1}{r+1} \frac{\sqrt{S^{(k-1)}_{1,r+1} S^{(k+1)}_{1,r+1}}}{S^{(k)}_{1,r}} 
\end{align}

In the perturbative regime, for $r \ll k$, this becomes 
\begin{align}
\langle \phi^{IR}_{r-1,r} \phi^{UV}_{r,r+1} | RG \rangle &=  \frac{\sqrt{r^2-1}}{r} \cr
\langle \phi^{IR}_{r+1,r} \phi^{UV}_{r,r-1} | RG \rangle &=  \frac{\sqrt{r^2-1}}{r} \cr
\langle \phi^{IR}_{r-1,r} \phi^{UV}_{r,r-1} | RG \rangle &= - \frac{1}{r}  \cr
\langle \phi^{IR}_{r+1,r} \phi^{UV}_{r,r+1} | RG \rangle &= \frac{1}{r} 
\end{align}

Comparing with the literature, we find perfect agreement with the linear combinations of UV fields 
which are supposed to give the IR fields:
\begin{align}
 \phi^{IR}_{r-1,r} =  \sqrt{r^2-1} \phi^{UV}_{r,r+1} - \phi^{UV}_{r,r-1} \cr
  \phi^{IR}_{r+1,r} = \phi^{UV}_{r,r+1} +  \sqrt{r^2-1}   \phi^{UV}_{r,r-1} 
\end{align}

At the next level of complexity, one can discuss the mixing of the fields $\phi^{UV}_{r,r\pm 2}$ and $\partial \bar \partial \phi^{UV}_{r,r}$
to give $\phi^{IR}_{r\pm 2,r}$ and $\partial \bar \partial \phi^{IR}_{r,r}$. This is a considerably more laborious task. 
We computed the desired one-point functions:
\begin{align}
\langle \phi^{IR}_{r+2,r} \phi^{UV}_{r,r+2}| RG \rangle &= \frac{2 k (k+4)-3 \left(r^2+r-2\right)}{(r+1)(r+2) (k-r+1) (k+r+3)}\frac{\sqrt{S^{(k-1)}_{1,r+2} S^{(k+1)}_{1,r+2}}}{S^{(k)}_{1,r}}  \cr
\langle \phi^{IR}_{r+2,r}  (2 h^{UV}_{r,r})^{-1} \partial \bar \partial \phi^{UV}_{r,r}| RG \rangle &= \frac{2 k (k+r+5)+5 r+11}{(k+2) (r+1) (k+r+3)} \frac{\sqrt{S^{(k-1)}_{1,r+2} S^{(k+1)}_{1,r}}}{S^{(k)}_{1,r}} \cr
\langle \phi^{IR}_{r+2,r} \phi^{UV}_{r,r-2} | RG \rangle &= \frac{\sqrt{S^{(k-1)}_{1,r+2} S^{(k+1)}_{1,r-2}}}{S^{(k)}_{1,r}} \cr
\langle \phi^{IR}_{r-2,r} \phi^{UV}_{r,r-2}| RG \rangle &= \frac{2 k (k+4)-3 (r-2) (r+1)}{(r-1)(r-2)(k-r+3) (k+r+1)} \frac{\sqrt{S^{(k-1)}_{1,r-2} S^{(k+1)}_{1,r-2}}}{S^{(k)}_{1,r}}  \cr
\langle \phi^{IR}_{r-2,r}  (2 h^{UV}_{r,r})^{-1} \partial \bar \partial \phi^{UV}_{r,r}| RG \rangle &=  -\frac{2 k (k-r+5)-5 r+11}{(k+2) (r-1) (k-r+3)} \frac{\sqrt{S^{(k-1)}_{1,r-2} S^{(k+1)}_{1,r}}}{S^{(k)}_{1,r}}  \cr
\langle \phi^{IR}_{r-2,r}  \phi^{UV}_{r,r+2}| RG \rangle &= \frac{\sqrt{S^{(k-1)}_{1,r-2} S^{(k+1)}_{1,r+2}}}{S^{(k)}_{1,r}}  \cr
\langle  (2 h^{IR}_{r,r})^{-1}\partial \bar \partial \phi^{IR}_{r,r}  \phi^{UV}_{r,r+2}| RG \rangle &= \frac{2 k (k-r+3)-3 r+3}{(k+2) (r+1) (k-r+1)} \frac{\sqrt{S^{(k-1)}_{1,r} S^{(k+1)}_{1,r+2}}}{S^{(k)}_{1,r}}  \cr
\langle  (2 h^{IR}_{r,r})^{-1} \partial \bar \partial\phi^{IR}_{r,r} (2 h^{UV}_{r,r})^{-1} \partial \bar \partial \phi^{UV}_{r,r} | RG \rangle &=\frac{k (k+4) \left(r^2-5\right)+2 \left(r^2-7\right)}{(k+2)^2 \left(r^2-1\right)} 
\frac{\sqrt{S^{(k-1)}_{1,r} S^{(k+1)}_{1,r}}}{S^{(k)}_{1,r}} \cr
\langle  (2 h^{IR}_{r,r})^{-1} \partial \bar \partial\phi^{IR}_{r,r} \phi^{UV}_{r,r-2}| RG \rangle &= \frac{-2 k (k+r+3)-3 (r+1)}{(k+2) (r-1) (k+r+1)} \frac{\sqrt{S^{(k-1)}_{1,r} S^{(k+1)}_{1,r-2}}}{S^{(k)}_{1,r}} 
\end{align}

From the conformal perturbation theory we expected at leading order
\begin{align}
\phi^{IR}_{r+2,r} &= \frac{2}{r(r+1)} \phi^{UV}_{r,r+2}+ \frac{2}{r+1} \sqrt{\frac{r+2}{r}} (2 h^{UV}_{r,r})^{-1} \partial \bar \partial \phi^{UV}_{r,r} + \frac{ \sqrt{r^2-4}}{r} \phi^{UV}_{r,r-2} \cr
\phi^{IR}_{r-2,r} &= \frac{\sqrt{r^2-4}}{r} \phi^{UV}_{r,r+2} - \frac{2}{r-1} \sqrt{\frac{r-2}{r}} (2 h^{UV}_{r,r})^{-1} \partial \bar \partial \phi^{UV}_{r,r} + \frac{2}{r(r-1)} \phi^{UV}_{r,r-2}\cr
 (2 h^{IR}_{r,r})^{-1} \partial \bar \partial \phi^{IR}_{r,r} &= \frac{2}{r+1} \sqrt{\frac{r+2}{r}} \phi^{UV}_{r,r+2}+\frac{r^2-5}{r^2-1} (2 h^{UV}_{r,r})^{-1} \partial \bar \partial \phi^{UV}_{r,r} - \frac{2}{r-1} \sqrt{\frac{r-2}{r}} \phi^{UV}_{r,r-2}
\end{align}
The agreement is perfect.

\subsection{Alternative candidates}
The other $Z_2$-twisted boundary conditions for $\CT_\CB$ can be obtained by acting with the $\CB$-topological 
defects $\CD^{\CB}_{t,d,1,s}$ on the basic boundary condition we used for our candidate. If we 
act with $\CI_1$ on such $(t,s)$ boundary conditions, we get the same result as if we had acted with the 
$\CD^{UV}_{1,s}$ and $\CD^{IR}_{t,1}$ defects on our RG domain wall candidate. 

Hence the only criterion we 
can use to distinguish the correct RG domain wall candidate among the 
various $(t,s)$ images is the comparison to the conformal perturbation theory
results. For $(t,s)$ small compared to $k$, 
the modified RG candidates would give one-point functions 
which differ only by an overall $ts$ factor. Although this is already probably sufficient to 
show that $s=t=1$ is the correct RG domain wall candidate, in order to 
really discriminate sharply among the various  $(t,s)$ images it would be nice to 
extend the perturbative calculations to $\phi^{IR}_{r,r} \phi^{UV}_{r,r}$, 
$\phi^{IR}_{r\pm 1,r} \phi^{UV}_{r ,r\pm 1}$, etc. for $r$ of order $k$. 

\subsection{A sign puzzle}
We would conclude our calculations with a small puzzle.
Let's compute the one-point function of $\phi^{IR}_{1,1} \phi^{UV}_{1,3}$.
The field $\phi^{IR}_{1,1} \phi^{UV}_{1,3}$ is a level $1/2$ $\tilde \CB$ descendant of 
the NS field $(1,3,3)$. Thus we need to find the linear combination of primaries 
 $\epsilon\phi_{1,3,3} $ and $\phi_{1,1,3} \equiv G_{-1/2} \phi_{1,3,3}$ which has zero conformal dimension for $T^{IR}$. 
 We can compute
 \begin{align}
 L_0^{IR} \epsilon\phi_{1,3,3} &=  \frac{k-1}{2k+4} \epsilon\phi_{1,3,3} + \frac{ \sqrt{(k+1)(k+3)} }{2k+4}G_{-1/2}\phi_{1,3,3}  \cr
 L_0^{IR} G_{-1/2} \phi_{1,3,3}&=   \frac{k+1}{2k+4} G_{-1/2} \phi_{1,3,3} + \frac{k-1}{(k+3)} \frac{ \sqrt{(k+1)(k+3)} }{2k+4}\epsilon \phi_{1,3,3} \end{align}
 The correct linear combination is thus 
 \begin{equation}
 \phi^{IR}_{1,1} \phi^{UV}_{1,3} =\sqrt{\frac{k+1}{2k}}  \epsilon\phi_{1,3,3} -\sqrt{\frac{k+3}{2k}}  G_{-1/2} \phi_{1,3,3}
 \end{equation}
 and the overlap with the $Z_2$ image is $- k^{-1}$. 
 Thus we have the one point function 
 \begin{align}
\langle \phi^{IR}_{1,1} \phi^{UV}_{1,3} | RG \rangle =  - \frac{\sqrt{3}}{k} 
\end{align}

This sign is somewhat disturbing, even as the factor of $\sqrt{3}$ is rather nice: the leading value of the coupling constant in the IR is expected to be 
\begin{equation}
2 \pi g_* \sim \frac{\sqrt{3}}{k}
\end{equation}
and the one-point function of $\phi^{IR}_{1,1} \phi^{UV}_{1,3}$ should be closely related to that. But the sign seems wrong: the critical coupling constant should be positive!
The sign would not vary if we were to use some other RG domain wall candidate, as long as $s, t$ are small compared to $k$.
It is possible we may have missed subtle overall signs in front of the Ishibashi state. The various topological defects and brane we used 
involve square roots of the modular matrix elements. 

\section{More general cosets}
There is a conjectural RG flow between the cosets
\begin{equation}
\CT_{UV} = \frac{\hat g_l \times \hat g_m}{\hat g_{l+m}} \qquad \qquad m > l
\end{equation}
and
\begin{equation}
\CT_{IR} = \frac{\hat g_l \times \hat g_{m-l}}{\hat g_{m}} \qquad \qquad m > l
\end{equation}
initiated by the $\phi=\phi_{1,1}^{\mathrm{Adj}}$ coset field \cite{Ravanini:1992fs}.

It should be clear that $\phi$ is still transparent to topological defects associated to representations 
$[r_1,r_2;1]$ of the UV coset, and that these will match the IR defects labeled by $[r_1,1;r_2]$. 
The same reasoning as for the minimal models suggests one to look at the 
coset theory  
\begin{equation}
\CT_{\CB} = \frac{\hat g_l \times \hat g_l \times \hat g_{m-l}}{\hat g_{l+m}} \qquad \qquad m > l
\end{equation}
We have again a natural $Z_2$ automorphism, and 
we can try to build the RG boundary condition as an image 
of a $Z_2$-twisted $\CT_\CB$ brane. 

We can even propose a natural candidate $\CT_\CB$ brane 
\begin{equation}
|\tilde B \rangle = \sum_{s,t} \sqrt{S^{(m-l)}_{1,t} S^{(m+l)}_{1,s}} \sum_d | t, d,d,s; \CB, Z_2 \rrangle 
\end{equation}

There is also a conjectural (never perturbative!) RG flow between the parafermionic theory \cite{Fateev:1991bv} 
\begin{equation}
\CT_{UV} = \frac{\hat{su}(2)_k}{\hat u(1)_k} 
\end{equation}
and the minimal model 
\begin{equation}
{\cal M}_{k+2,k+1} = \frac{\hat {su}(2)_{k-1} \times \hat {su}(2)_1}{\hat {su}(2)_{k}}
\end{equation}
initiated by perturbation by the basic parafermionic fields. 

The coset structure suggests this may be another example amenable of analysis by our methods,
with  
\begin{equation}
\CT_{\CB} = \frac{\hat{su}(2)_{k-1} \times \hat {su}(2)_1 }{\hat u(1)_k} 
\end{equation}
It would be interesting to identify an appropriate class of twisted boundary conditions. 

\acknowledgments{The work of DG is supported in part by NSF grant NSF PHY-0969448
and in part by the Roger Dashen membership in the Institute for Advanced Study.
Opinions and conclusions expressed here are those of the authors and do not necessarily reflect the views of funding agencies.} \\

\bibliographystyle{JHEP}
\bibliography{2dcft}

\end{document}